\documentclass[conference]{IEEEtran}
\usepackage{url}
\usepackage[pdftex]{graphicx}
\pdfoutput=1 

\usepackage[margin=1.5cm]{geometry}
\usepackage{array, xcolor}
\usepackage{lipsum}
\usepackage{bibentry}
\usepackage{longtable}
\usepackage{authblk}
\usepackage{times}
\usepackage{helvet}
\usepackage{courier}
\usepackage{color}
\usepackage{soul}
\usepackage{url}
\usepackage{xcolor}
\usepackage{stfloats}
\usepackage{caption}
\usepackage{tabularx}
\usepackage{epstopdf}
\usepackage{rotating}
\usepackage{epstopdf}
\usepackage{siunitx}
\usepackage{booktabs}
\usepackage{subcaption}
\usepackage{tabularx}
\usepackage{subcaption}
\usepackage{array}

\begin{document}

%\title{Taming Uber's Surge Pricing Mechanics With Data Driven Mobile Apps}
\title{Mining Open Datasets for Transparency in Taxi Transport in Metropolitan Environments}

\author[1]{Anastasios Noulas}
\author[2]{Vsevolod Salnikov}
\author[2]{Renaud Lambiotte}
\author[1]{Cecilia Mascolo}

\affil[1]{Computer Laboratory, University of Cambridge, UK}
\affil[2]{naXys, University of Namur, Belgium}

\maketitle

Uber has recently been introducing novel practices in urban taxi transport. 
Journey prices can change dynamically in almost real time and also vary geographically from one area to another in a city, a strategy known as surge pricing. 
In this paper, we explore the power of the new generation of open datasets 
towards understanding the impact of the new disruption technologies that emerge in the area of
public transport. With our primary goal being a more transparent economic landscape for urban commuters, we provide a direct price comparison between Uber and the Yellow Cab company in New York. 
We discover that Uber, despite its lower standard pricing rates, effectively charges higher fares
on average, especially during short in length, but frequent in occurrence, taxi journeys. 
Building on this insight, we develop a smartphone application, \textit{OpenStreetCab}, that offers a personalized consultation to mobile users on which taxi provider is cheaper for their journey. Almost five months
after its launch, the app has attracted more than three thousand users in a single city. Their journey queries have
provided additional insights on the potential savings similar technologies can have for urban commuters, with a highlight being that on average, a user in New York saves 6 U.S. Dollars per taxi journey if they pick the cheapest taxi provider. We run extensive experiments to show how Uber's surge pricing is the driving factor of higher journey prices and therefore higher potential savings 
for our application's users. Finally, motivated by the observation that Uber's 
surge pricing is occurring more frequently that intuitively expected, we formulate a prediction task
where the aim becomes to predict a geographic area's tendency to surge. Using exogenous to Uber data, in particular Yellow Cab and Foursquare data, we show how it is possible to estimate customer
demand within an area, and by extension surge pricing, with high accuracy.

\section{introduction}
\label{sec:intro}
The arrival of Uber~\cite{uber} and its growing popularity have introduced an unprecedented change in the nature of taxi transportation: \textit{Pricing patterns can now change in every coming minute, driven by algorithmic recipes based on offer and demand put forward by the company. In addition, recent empirical findings ~\cite{wpostsurge} demonstrated that Uber's changes in pricing, a tactic popularly known as surge pricing, can vary from one neighborhood to the next one in a city.}
This situation translates into an extremely volatile pricing landscape in taxi transport, with prices changing in real time in a manner that is hard to predict or trace. Moreover, the precise working of pricing algorithms  is neither known to the public nor to authorities. As a result, the a-priori knowledge and transparency on pricing in urban transport, which has been a norm for decades, is effectively lost. 

In recent years, data mining research has focused primarily on the mining of spatial trajectories for the development of routing, navigation and mapping applications~\cite{giannotti2007trajectory,
bogorny2010spatial, o2014mining}. 
While taxi spatial trajectory data has also been exploited heavily in this context~\cite{yuan2010t, zheng2011computing, zheng2011urban}, there is only little work on the mining of taxi mobility data in the light of other layers of data
and in particular those that can provide valuable information on the economic costs of taxi journeys. This could be attributed to the relatively stable prices in the taxi
industry for years now, but also to the existence of clear rules determining the price of a trip based on its duration and distance. The case of Uber as a game changer in urban transport economics  has motivated us to consider taxi mobility data from an economical point of view, in order to estimate and compare the financial costs incurred by customers of different taxi providers. Our goal here is set to answer a number of research questions that concern the relationship between taxi mobility patterns and the financial impact of those through the comparison of taxi providers over time and across space. 

En route to this goal, whose achievement is a first step to restore transparency for commuters in taxi transport, we make the following contributions in the present paper. 

\begin{itemize}
\item First, we leverage on a large, free and open dataset of yellow taxi
cab mobility records in New York City to characterize their mobility and pricing patterns. We report 
that pricing directly relates to well known patterns observed in the past on human urban mobility. Most taxi movements are within a short distance range with longer movements occurring less frequently in the data.
Further, the overall distribution of spatial movements directly matches the statistical distribution of the taxi fares paid by customers. This observation is due to the inherent relationship between the magnitude of mobility trajectories and their financial or energy costs. 
Next, we provide a head to head comparison of two taxi providers competing in New York City: yellow cabs and Uber's cheapest service, Uber X. We note that, while the statistical distributions of prices charged
between the two companies follows a similar pattern, Uber X appears to be consistently more expensive on average.
In particular, Uber takes effectively advantage of trends in human mobility patterns, charging more for short trips and thus maintaining a higher revenue margin (Section~\ref{sec:analysis}). 
\item We take a step further and build a mobile application, \textit{OpenStreetCab}\footnote{\url{www.openstreetcab.com}}, that allows users to query the origin (pick up) and destination (taxi drop off) locations of their journey. The more than three thousand users that have used the application in New York city have generated thousands of mobility and pricing datapoints that have allowed us to perform an additional data mining step that reveals the large potential benefits of big open datasets in the context of urban transport. Specifically, taxi commuters that use the app save on 
average an estimated amount of 6 U.S. Dollars per journey. A deeper inspection of the data demonstrates that savings, as driven by the surge pricing patterns imposed by Uber, can vary significantly by the hour of the week and by user location (Sections~\ref{sec:app}~and~\ref{sec:evaluation}).
\item While the findings initially appear to be in contradiction with the standard pricing reported by Uber, we discover that higher prices - compared to the publically stated base fares - are being charged very frequently (almost one in four times). For this reason, the effective price incurred on taxi customers is higher than 
the stated and expected minimum. We perform two controlled experiments aiming to reverse engineer the surge pricing tactics of Uber. 
We show that surge pricing is enabled very frequently, with per minute sensitivity, based on supply and demand balance at the origin and also, possibly, at destination. Moreover, we demonstrate that surge pricing
has spatial structure and we exploit Yellow Cab and Foursquare data to predict demand at an area of a city, 
and by extension its tendency to surge (Section~\ref{sec:surge}). 
\end{itemize}
Overall, our work shows how the combination of open datasets and data generated by mobile applications can allow researchers and practitioners alike to understand complex phenomena in the urban domain. 
The rest of the paper is structured as follows. In Section~\ref{sec:analysis} we analyse the taxi mobility and fares datasets, where we provide a direct comparison between Uber X and Yellow Cabs. In Section~\ref{sec:app} we describe our application, OpenStreetCab, that leverages on these datasets to help commuters choose
the cheapest taxi provider for their journey. In Section~\ref{sec:evaluation}, we perform an analysis
on the data yielded by the app focusing on the savings made by mobile users, whereas in Section~\ref{sec:surge} we describe the surge pricing mechanics of Uber. Finally, we close with related works (Section~\ref{sec:related}) and concluding remarks (Section~\ref{sec:discussion}).

\section{analysis}
\label{sec:analysis}
In this Section we provide an overview of the dataset describing taxi mobility and 
fares charged in New York.  
We then evaluate the prices that Uber X would charge for trips sampled from the dataset and compare them with those charged by Yellow Cabs, considering aggregate, temporal and spatial comparative perspectives. 

\begin{figure}
  \centering
    \includegraphics[width=.7\linewidth]{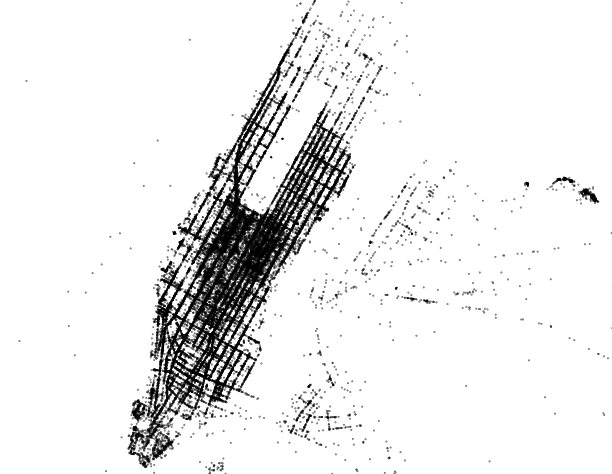}
\caption{Marking the traces of new york city yellow taxis. For every pick up and drop off point in a uniform sample of the data we draw a black point.}  
\label{taxitraces}  
\end{figure}

\paragraph*{\textbf{The New York City Taxi Dataset}}
The Freedom of Information Law in the United States encourages public authorities 
to release their data where appropriate to the benefit of the citizens. In 2014, 
the law was exploited by Chris Whong to acquire and post on the web one of the most 
comprehensive taxi mobility datasets available today. The dataset describes
taxi journeys in New York City during the full course of 2013, and informs us not only on the
origin and destination points of taxi trips in terms of geographic latitude and longitude coordinates,
but also on the financial costs for the customer (trip fare paid including information on tip amount and payment method). This mobility dataset,  downloadable here~\cite{Foiling}, counts 11GB of mobility 
data representing almost 170 million trips and 7.7GB of the associated fare data. 
Traces generated
by the data can be seen in Figure~\ref{taxitraces}, where we have drawn a black point for every pick up
and drop off point of a taxi journey considering a $1\%$ sample during January 2013 in the data. 

\paragraph*{\textbf{Comparing Prices between Taxi Providers}}
In August 2014, Uber opened up an API with access to valuable information about its services.
This occasion allowed  us to perform a first head to head comparative analysis of prices between Uber and Yellow taxis 
in New York City. To achieve this, we have run the following experiment during a 10 day time window in September 2014: 
\begin{enumerate}
\item  For a sample of $600$K trips in New York in the Yellow Taxi dataset, record the geographic coordinates (latitude and longitude) of the pick up and drop off points. 
\item  Retrieve the total fare paid by the customer for the trip (tip amount included).
\item  Query Uber's API on the corresponding endpoint and ask how much they would charge for the same trip (same pick up and drop off points), 
considering the cheapest version of the service, Uber X.
\item Uber's API returns a value range indicating the minimum and maximum price estimate. We take the mean of the two values.
\item  We then compare the prices between the two services and retrieve their difference. 
\end{enumerate}

\begin{figure}
  \centering
    \includegraphics[width=.9\linewidth]{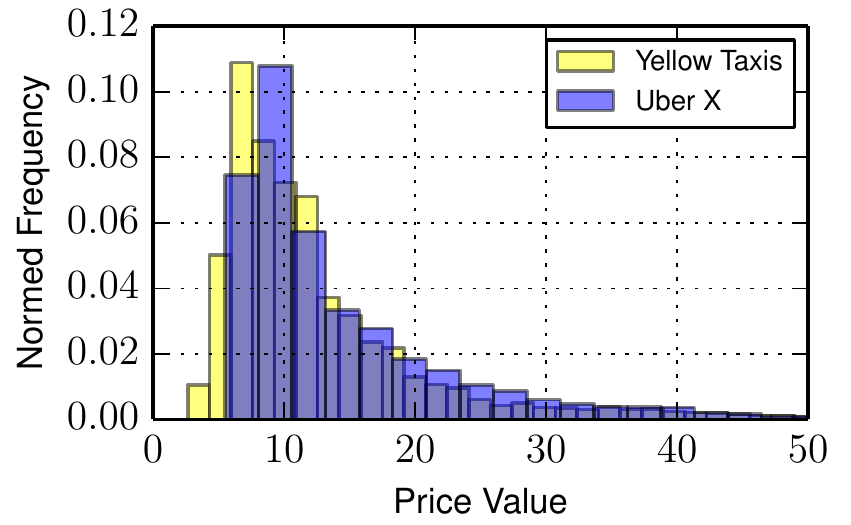}
\caption{Distribution of prices per journey for Uber X and Yellow Taxis in New York City.}  
\label{pricecomp}  
\end{figure}

\begin{figure}
  \centering
    \includegraphics[width=0.9\linewidth]{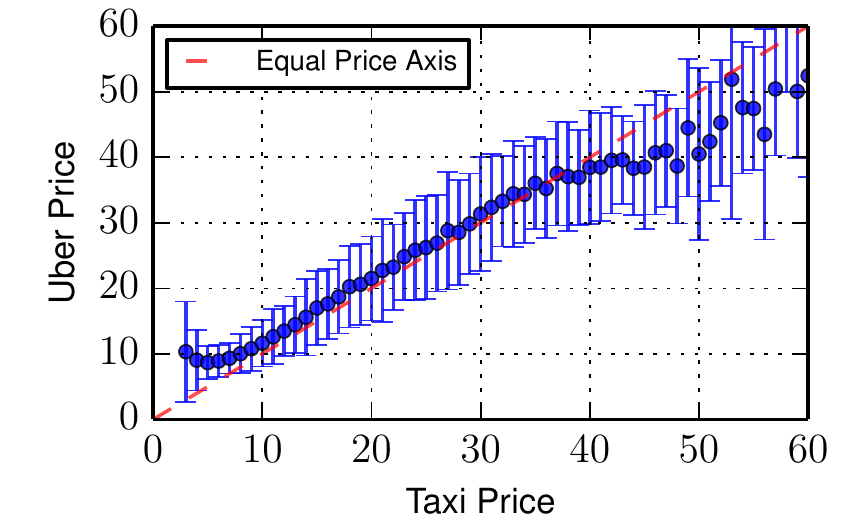}
\caption{Median Uber X price for a given Yellow Taxi price. Errors bars show one standard deviation from the average value.}  
\label{ubervsyellow}  
\end{figure}

\begin{figure}
  \centering
    \includegraphics[width=0.9\linewidth]{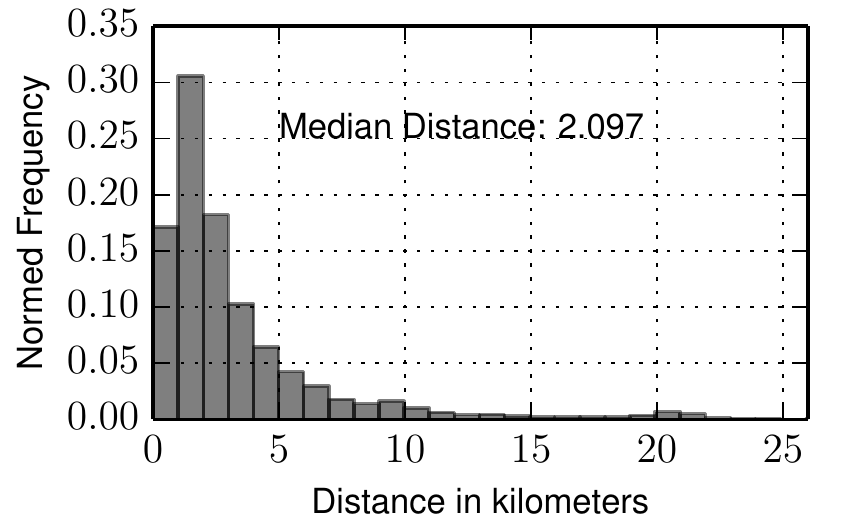}
\caption{Distribution of geographic distances between drop off and pick up points for Yellow Taxi journeys.}  
\label{tripdistances}  
\end{figure}

As can be observed in Figure~\ref{pricecomp} where the distribution of prices for the two services is shown, despite their qualitative similarity, yellow taxis appear on average (median) $1.4$ U.S. Dollars cheaper than Uber X. In Figure~\ref{ubervsyellow}, we compare Uber and yellow cabs from another perspective: for every observed yellow taxi price, we show the median Uber X price (one standard deviation noted through the error bars). 
If the two taxi service providers  cost the same for every trip, then a balanced relationship
would be found on the $x=y$ axis.
However, Uber appears consistently more expensive for prices below $35$ U.S.Dollars, becoming cheaper only above that threshold. As one would expect, the cheaper journeys are those that are in principle of shorter range. 
In fact, according to observations made on a variety of empirical data in the past, human mobility tends to be characterised by a vast majority of short trips~\cite{gonzalez2008understanding, brockmann2006scaling}, with a few, occasional very long ones. This observation suggests that Uber's economic model effectively exploits this trend of human mobility in order to maximise revenues. We empirically confirm this hypothesis noting the skewed frequency distribution of movement distances in the present context by visualising it in Figure~\ref{tripdistances}, where we measure a mean distance for a yellow taxi trip in New York equal to $2.09$ kilometers. The percentage of yellow
taxi journeys that cost less than $35$ U.S. Dollars is almost $94\%$.

\begin{figure}
  \centering
    \includegraphics[width=0.9\linewidth]{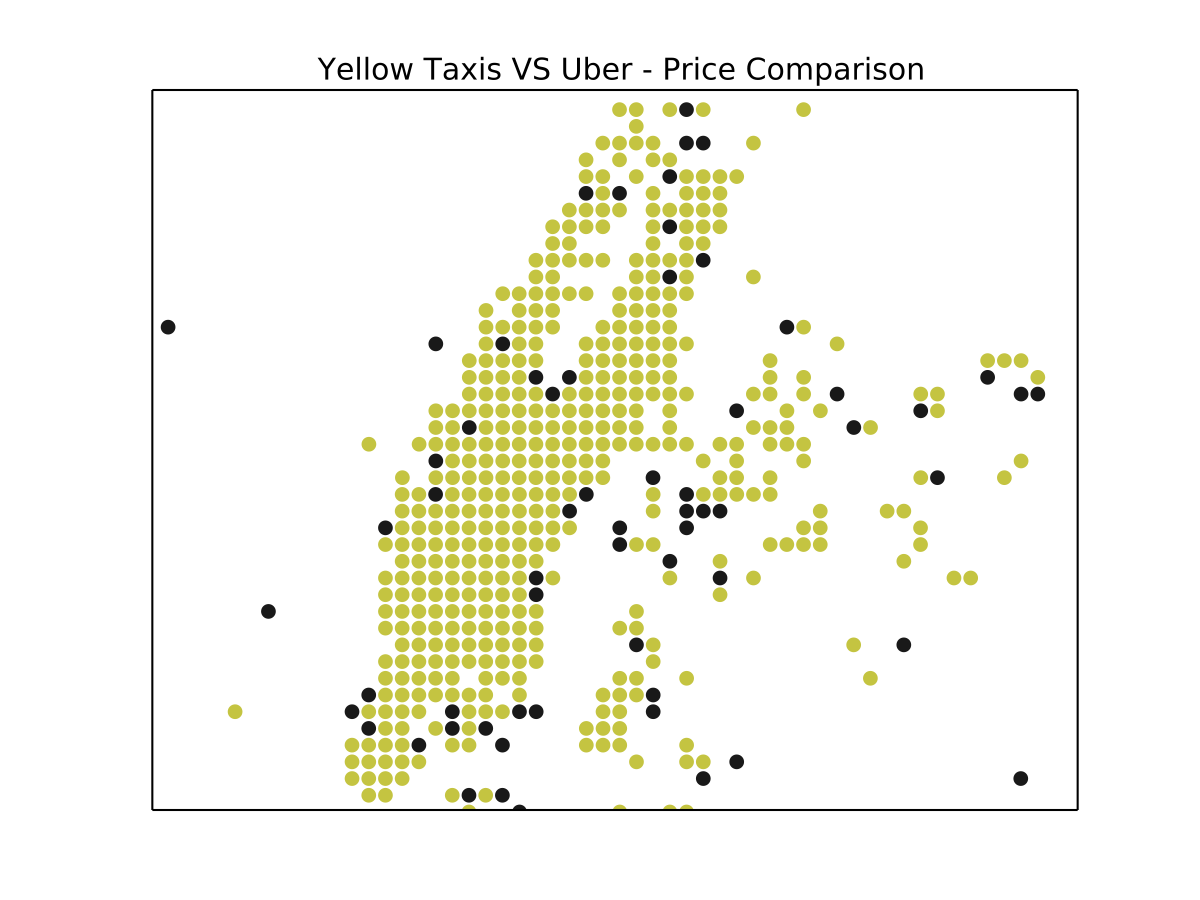}
\caption{Geographic comparison between Uber and Yellow Taxi prices. We paint an area black if Uber is cheaper by trip majority and yellow otherwise.}  
\label{geocomp}  
\end{figure}

%%% COMMENT ON MAP FIGURE %%%
In Figure~\ref{geocomp}, we put a geographic perspective on the comparison of the two taxi companies.
We split New York City in a set of grid areas ($100 \times 100$ meters). Considering then the set of all out-going trips from an origin area, we paint a given area yellow if most trips were cheaper when
taking a yellow cab. Instead, an area is painted black if Uber is cheaper by trip majority. 
One notes how the Manhattan area is typically cheaper for yellow taxis, confirming this area as an economic stronghold of the company~\footnote{A taxi medalion (licence) for the company costs $805$K U.S. Dollars as of 2015.}, whereas Uber is cheaper with higher frequency in the peripheral parts of the city. Since Uber considers the balance between driver supply and customer demand as factors to determine pricing~\cite{uberPricing}, it may be a plausible hypothesis that prices will be in general higher where there is high demand - that is the center of the city where population density surges - and at the same time where there is low driver supply. Supply may be prone to a geographic bias due to spatial variations in resident demographics. Most Uber drivers may not
reside in the very expensive Manhattan area and for this reason this area is likely to be more prone to surge pricing. 
%%% %%% %%%

The above experiment may involve a number of biases and limitations which we refer to here.
The NYC Yellow taxi data corresponded to year 2013 whereas our API requests for Uber X prices were made in September 2014. However, one should note that the prices for yellow taxis in the city had last changed in 2012 after 8 years~\cite{fares}. For this reason, prices in 2013 are expected to offer a good approximation of today's prices as, to the best of our knowledge, there has been no increase since 2012. Further, there was no control for time of the day/week for the API query, an additional dimension which should be incorporated when available. In particular, temporal information is expected to help predict variations of traffic, but also of offer and demand, and therefore of prices. Let us note, however, that surge pricing does not seem to be purely periodic, in terms of daily or weekly cycles, as we show in Section~\ref{sec:surge}. As more and more data is acquired, this temporal information could be incorporated into the analysis.  Preliminary analysis shows that repeating the same experiment at different time windows yields only minor changes in the numerical estimates presented above.

Overall, we argue that the comparison of two different companies providing the same service in the same geographic area is valuable to commuters. Just as consumers have had open access to airfares for a long time now, allowing for transparency in a  competitive market, we believe that similar approaches could benefit commuters in modern cities. For this reason, we design a mobile application that realizes this vision, as described in the next section. 

% \section{mobile application}
\section{OpenStreetCab: A mobile app for cheap taxi fare discovery}
\label{sec:app}
\begin{figure}
  \centering
    \includegraphics[width=1.0\linewidth]{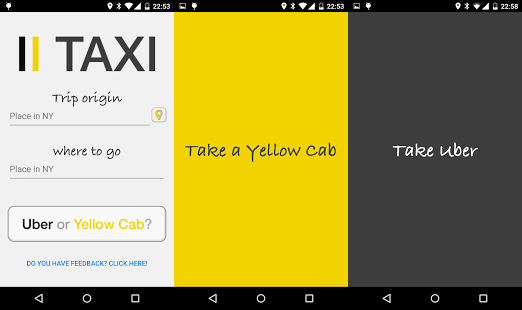}
\caption{A user perspective of OpenStreetCab. As shown in the snapshots above users can set their trip and destination address as they open the application. By pressing a button they receive a consultation on the cheapest taxi provider for their trip.}  
\label{appsnap}  
\end{figure} 

% \begin{figure}
%  \centering
%  \includegraphics[width=6.0cm]{images/gridAppLogic.pdf}
%  \caption{}  
% \label{applogic}  
% \end{figure}

% \begin{figure*}
%  \centering
%  \includegraphics[width=18.0cm]{images/weeklyWinners.pdf}
%  \caption{A snapshot of 168 hours in a week, coloured yellow or black depending on whether a yellow cab or an Uber offered the cheapest deal.}  
% \label{weeklycomparison}  
% \end{figure*}

In recent years, mobile applications have often be used as a source of data. Smartphones are pervasive devices following users through their daily activities, sensing their whereabouts and context. The corresponding data has fueled a number of studies, and led to the improvement and creation of many real world applications. Our analysis in the previous section shows that the price of a journey can  significantly vary from one provider to another, and that this variation is associated to the duration of the trips, as well as on where they take place. 
Motivated by these observations we have taken a step forward by designing and launching a mobile application, \textit{OpenStreetCab}, whose aim is to help users reduce commuting costs by taxi. 
This is achieved by helping users chose the cheapest taxi provider depending on the parameters of their journey. In this section, we first summarise the ideas behind the design  and functionality of the application. Next we show how the dataset generated through the app can also yield valuable insight on taxi economics, focusing on savings made by mobile users.

\paragraph*{\textbf{Application logic and functionality}}

Figure~\ref{appsnap} shows three snapshots
of the Android version of the app (iOS one is available as well). Users can provide as input their pick up (origin) and drop off (destination) locations. After clicking on  the button \textit{Uber or Yellow Cab?}, the query input is pushed to a server where Uber and yellow taxi prices are compared. If Uber X is found to be cheaper, on average, for the selected trip  a black screen is shown on the phone of the user with the message \textit{Take Uber}. Otherwise, if a yellow cab is cheaper for that journey, the screen becomes yellow with the message \textit{Take a Yellow Cab}. Minimalism in design is central to provide the user with an answer with a minimum cost in terms of actions.

The decision of whether  Uber X or yellow cab is cheaper is the most critical part of the application. 
We now describe how we use data from yellow taxi and Uber in New York (as discussed in Section~\ref{sec:analysis}) and Uber, and how the decision-making algorithm behind the service is built. 
\begin{enumerate}
\item First, we apply a grid on top of New York's geographic landscape. Its size is 400 by 400 number of cells, 
and each cell has size $30 meters \times 30 meters$.
\item The origin and destination  input by the user are geo-coded to latitude and longitude geographic coordinates. 
\item The coordinates are subsequently matched to their corresponding grid cells, denoted by $O$ for the origin and $D$ for the destination. 
\item We  calculate the yellow cab price, by taking the mean price across all journeys  starting in the origin cell $O$ and finishing in the destination cell $D$. The tip is taken into account in the price.
\item We query the Uber API in real time with, as an input, the geo-coded origin and destination addresses provided by the user. Uber returns a [min, max] estimate for Uber X and we consider its mean as the price of the trip.   
\item We compare Uber X against the Yellow Cab price and declare as winner the cheapest provider.
\end{enumerate}

With regard to step 4, a crucial aspect was to find the right level of granularity, not too coarse to avoid washing out useful signals, nor too narrow to avoid having a limited number of occurrences for the trips selected by the user. For instance, we have considered the possibility to stratify the historic journeys of yellow cabs by time. At different hours of the week, yellow cab prices may change due to difference in traffic conditions or commuting patterns. External phenomena such as weather conditions or large events can also have an
effect on the duration of a taxi journey. However, stratifying by time leads to less data per area and, as a consequence, worse estimates. For this reason, we have opted for a simple averaging of the prices for journeys that falls between the origin and destination cells. We have instead kept the cell size as small as possible, to $0.0009 km^2$ ($30m \times 30m$), to emulate the size of a small block in the city and  be as precise as possible  geographically.

% Our observations in the previous section show that it might be financially advantageous for taxi travellers to chose either a Yellow Cab or Uber, depending on the whereabouts of their journey. 
% In order to help users to take the right decision, we have developed a smartphone app, called OpenStreetCab, designed as follows.

% One limitation for the design of our service is that only prices for trips with origins and destinations 
% in the New York City Taxi Dataset can in principle be retrieved. In order to evaluate the price of any trip, as needed for a usable App, we have divided the NY region into a mesh with cells of size around 100m by 100m in order to index trips in the database efficiently. For each user query, we find a set of trips in our dataset with the origin in neighbouring cells of desired origin and, among them, we find the trip whose destination is closest to the desired one. This strategy has the advantage of being sufficiently fast to perform online queries and expected to provide reliable price estimates. For the same trip, Uber price is obtained through their API.

\section{analysis of potential savings}
\label{sec:evaluation}
\paragraph*{\textbf{Basic Data Properties and Analysis}}
\textit{OpenStreetCab} was launched in March 2015 and in less than three months has been installed by more than $4.5$K iPhone and Android users only in New York. In the latest app version, users are not only informed of the  cheapest taxi provider for their journey, but also how much they would save in U.S. Dollars with the optimal choice. At least $3.5$K users have used the app at least once with the total number of queries being around $6.0$K. The average number of queries per user is $3.3$. 

\begin{figure}
\centering
\includegraphics[width=1.0\linewidth]{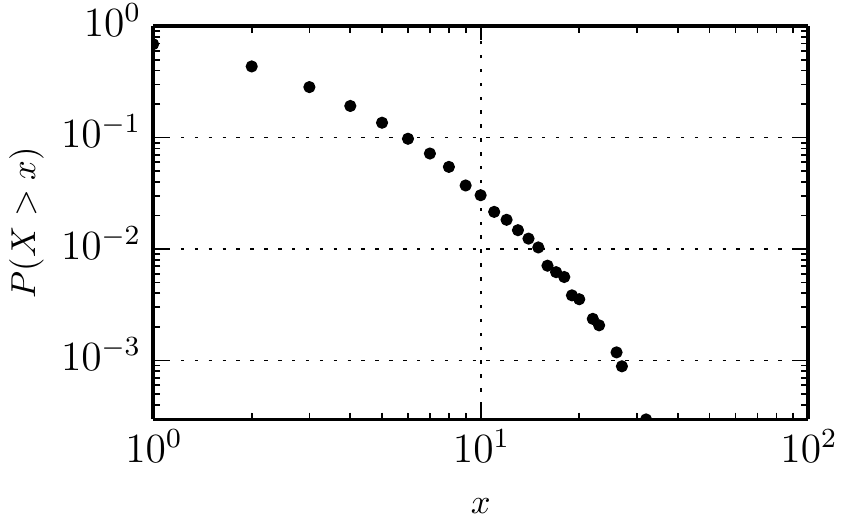}
\caption{Cumulative Distribution Function of Queries.}  
\label{queryDistr}  
\end{figure}

In Figure~\ref{queryDistr} we plot the Cumulative Distribution Function (CDF) of user query frequencies. The CDF follows a fat-tailed  distribution with the majority of users having queried the application only a few times and a few active users having used
the app several times. $10$\% of users have used the app more than 7 or 8 times, and 
a few handful of them ($1-2$\%) have queried the app more than 15 times so far. 
The usage statistics present an expected long tail, as observed in a variety of social datasets, including the number of phone calls placed by a person and, therefore, its number of geographic localisation in Call Detail Records data~\cite{gonzalez2008understanding}. 

\begin{figure}
\centering
\includegraphics[width=1.0\linewidth]{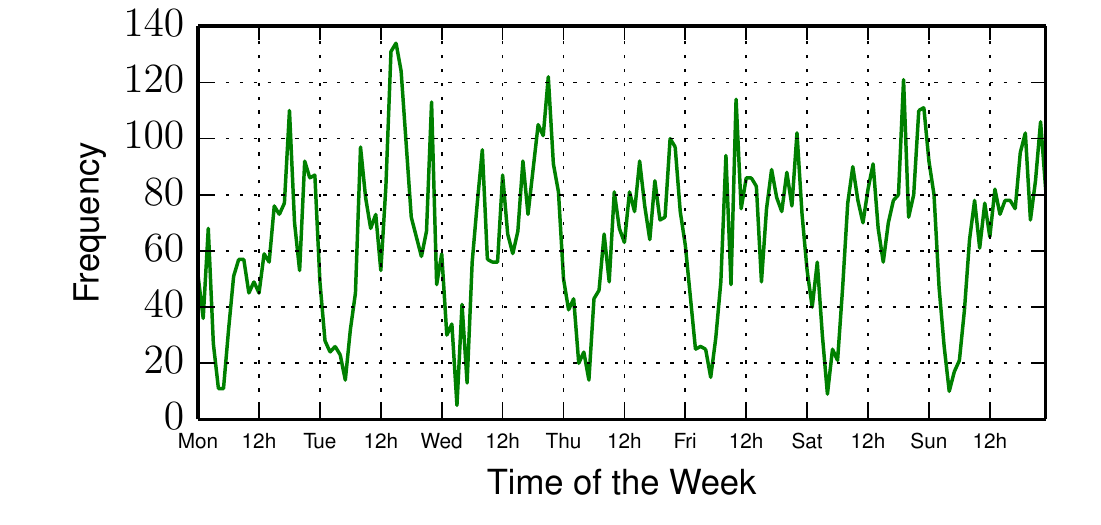}
\caption{User Query Frequency in terms of weekly temporal evolution patterns.}  
\label{queryWeekly}  
\end{figure}

\begin{figure}
\centering
\includegraphics[width=1.0\linewidth]{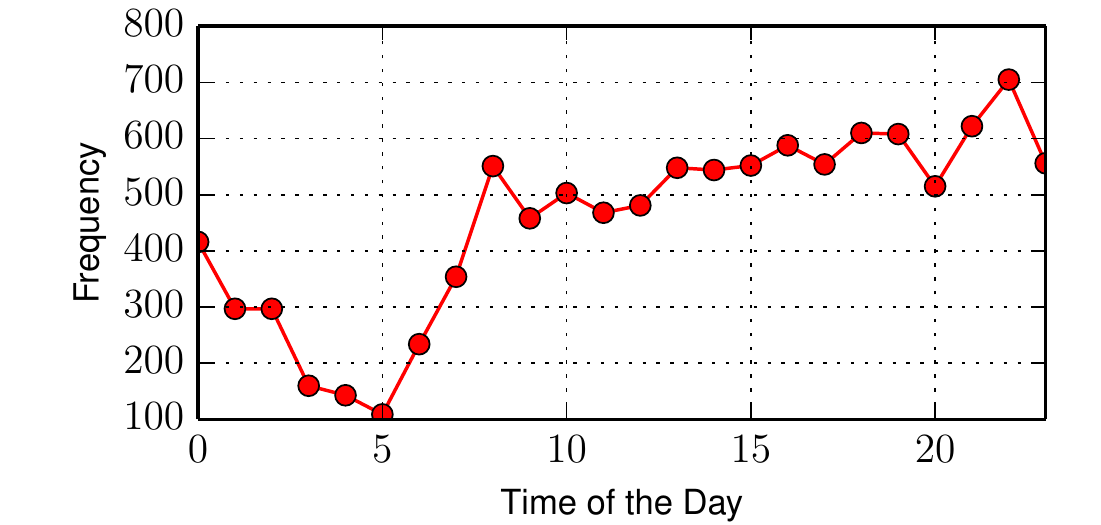}
\caption{User Query Frequency in terms of daily temporal evolution patterns.}  
\label{queryDaily}  
\end{figure}

In Figure~\ref{queryWeekly} we plot the weekly frequency of travel queries made to the app. The primary observation lies on the fact that Tuesday to Saturday are the most active days in terms of user engagement. Secondly, during the interval of a day (24 hours), we observe two characteristic peaks: a sudden rise in activity in the morning corresponding to early day commuters and a second one late in the evening when people return home. Note that our user base 
is inherently formed by Uber users in New York. Figure~\ref{queryDaily} shows the 24-hour frequency distribution of queries, averaging across all days, and confirms these 
observations.

\paragraph*{\textbf{User Savings on Taxi Transport}}

Let us now estimate the savings generated by our app. Considering $10,873$ 
travel queries in total, we iterate through the full set of query records 
and measure how much a user saves by taking the absolute difference in the prices 
between the two taxi providers. Formally for a queried journey $i$, we note 
the price difference, $\Delta \tau_{i} $ equal to $Yellow(\tau_{i})- Uber(\tau_{i})$.

\begin{figure}
\centering
\includegraphics[width=1.0\linewidth]{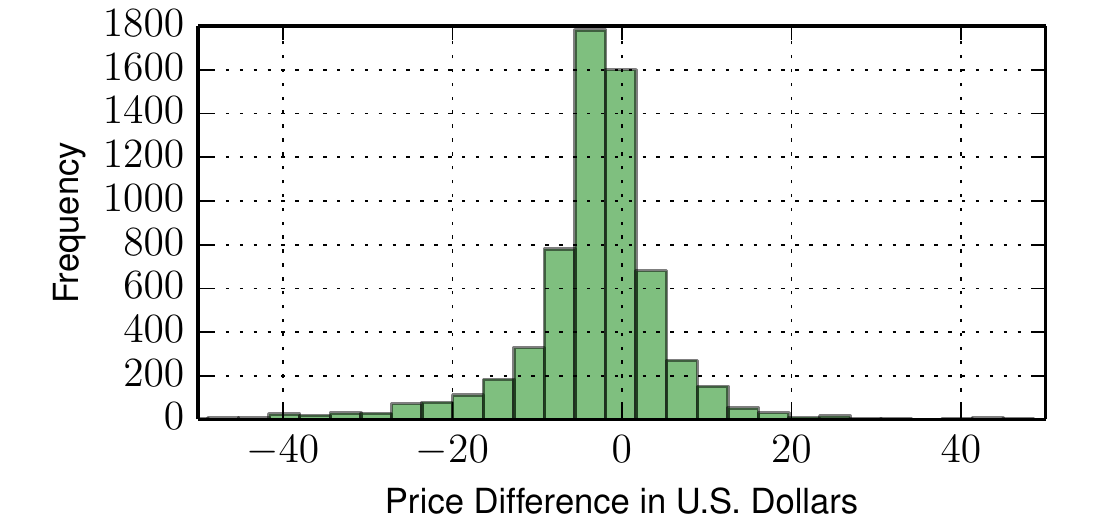}
\caption{Distribution of price differences (Yellow Taxi Price - Uber X Price) for all journeys queried through the app.}  
\label{Savings}  
\end{figure}

In Figure~\ref{Savings} we plot the histogram of $\Delta \tau$ considering all journeys. A difference of $0$ indicates that, based on our estimations, the two  providers charge the same amount 
for the journey requested by the user. The distribution is centred 
around  zero, but it exhibits a large variance, which translates into substantial potential savings for the users. We have measured 
an average saving per  journey equal to $6.05$ U.S. Dollars. This number should be put in perspective with the observation that most trips fall in the cost range $(7-15)$ U.S. Dollars, thereby indicating that important savings could be made by properly estimating and comparing the prices of competing operators. 

\begin{figure*}
 \centering
 \includegraphics[width=18.0cm]{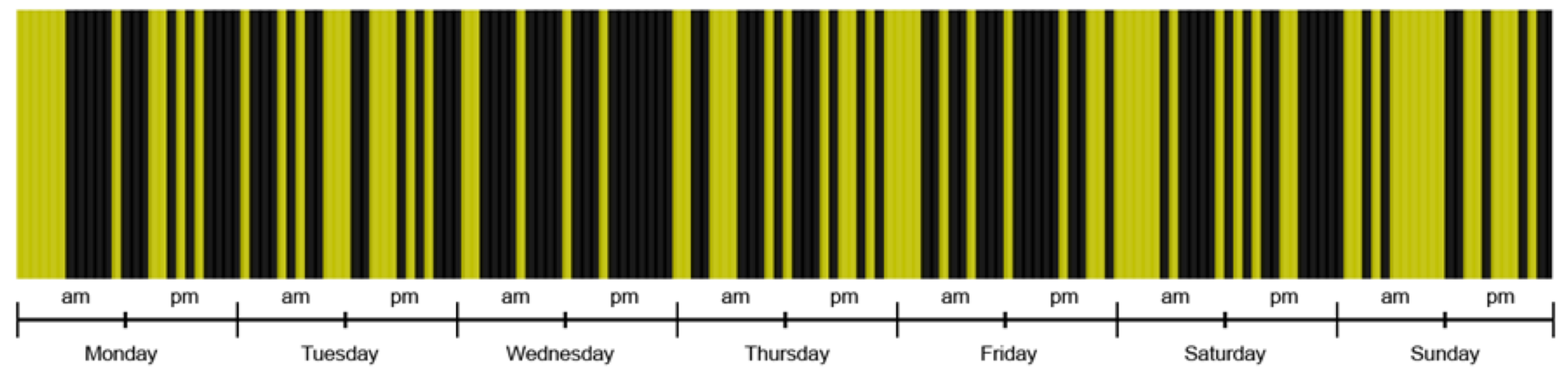}
 \caption{A snapshot of 168 hours in a week, coloured yellow or black depending on whether a yellow cab or an Uber offered the lowest price.}  
\label{weeklycomparison}  
\end{figure*}

Does \textit{when} help choosing the cheapest taxi provider in the city? In Figure~\ref{weeklycomparison}, each hour of the week has been coloured in a yellow or black stripe, depending on whether the majority of Uber or yellow cabs rides were cheaper for the hour in question. The visualization suggests that the time of the week can play a significant role in pricing. Interestingly, this temporal pattern is not purely periodic, as it depends on variations in traffic and on Uber's pricing model, itself depending dynamically on driver supply and customer demand. This preliminary observation, which demands further analysis, shows that, depending on the time of the week, it could be beneficial to pick one provider or another. 

% This motivates the future development of a real-time 
% application informing taxi users on their taxi options and corresponding financial costs. 

%%% saving strategies paragraph %%%
Finally, to provide a deeper insight on how different taxi pick-up strategies can be more
or less financially beneficial  for a user, we consider the following experiment. Running through all travel queries in the app's database we measure the 
cost $c_{ij}$ of a trip $i$ when using a given pick up strategy $j$. We consider four pick up strategies as described below: 
\begin{enumerate}
\item \textit{Application-driven}: The user  always takes the cheapest provider according to the output provided by OpenStreetCab.
\item \textit{Always Yellow Cab}: The user  always picks a yellow cab ignoring the app's output. 
\item \textit{Always Uber X}: The user always picks a Uber X driver for their journey.
\item \textit{Random Pick Up}: The user picks a taxi provider at random.
\end{enumerate}
%%%%%%%%%%%%%%%%%%%%%%%%%%%%%%%%%%

\begin{figure}
\centering
\includegraphics[width=1.0\linewidth]{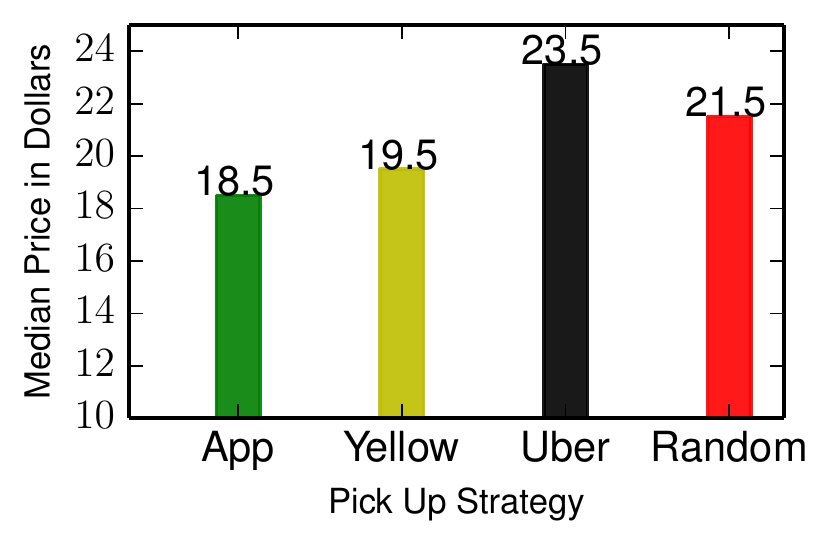}
\caption{Saving Strategies Median Prices considering four different strategies that could be 
hypothetically followed by our app's user base.}  
\label{SavingStrategies}  
\end{figure}

In Figure~\ref{SavingStrategies} we show the average savings obtained for each of
the strategies defined above. The application-driven strategy suggests a mean price of $18.5$
U.S. Dollars, when the next optimal strategy appears to be the one that always suggests
taking a yellow cab ($19.5$). Interestingly, taking Uber always is worse even than a random pick up strategy. This contradicts the  low cost image advertised by Uber
based on their own ratings, in part because of the large prevalence of short trips where yellow cabs was shown to be advantageous, but also because of the so-called surge pricing. For this reason, we explore in the next section the spatial and dynamical properties of Uber's pricing strategy.

\section{surge pricing}
\label{sec:surge}
The analysis in the previous sections shows how Uber introduces 
a new economic paradigm in the area of urban transport. 
The spearhead of this transformation is the surge pricing tactics 
enforced by the algorithmic recipes of the company. As we have observed already, 
taxi journey prices can 
vary in real time and from one neighborhood to another. Moreover the variations
can have significant implications on the costs incurred on travellers. 
Motivated by these observations we consider the following questions in this section:
First, \textit{How does surge pricing manifest in the city over time and space?} and second,
\textit{Can we exploit different data sources to predict Uber's surge pricing
patterns?}

\paragraph*{\textbf{Surge Pricing Patterns}}

\begin{figure}
\centering
\includegraphics[width=1.0\linewidth]{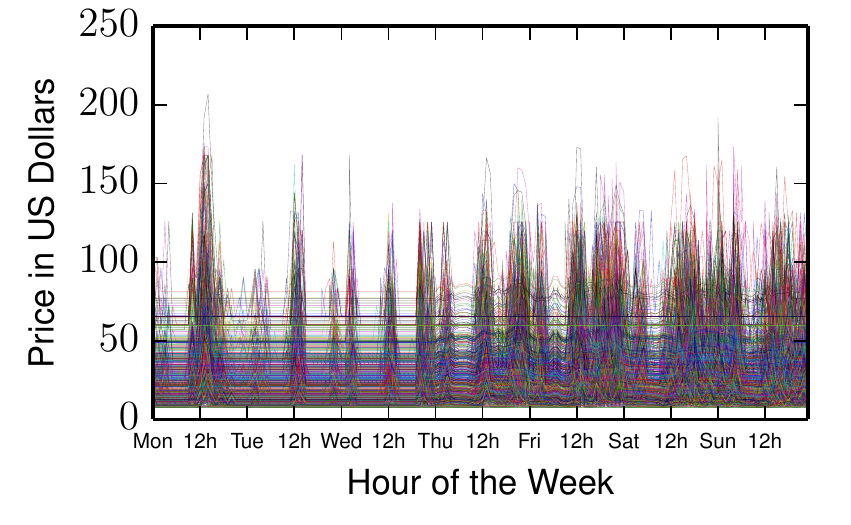}
\caption{Price Evolution Temporal Dynamics for a set of $800$ routes that where sampled 
uniformly random by our app's set of requested routes. The price of each route has been
queried once every hour for a week in April 2015.}  
\label{priceEvol}  
\end{figure}

In Figure~\ref{priceEvol} we plot the temporal variation of prices for a sample of $800$ routes queried by our app's users. Each drawn curve corresponds to the price of a route over time, with the price noted  on the y-axis. We have used a sampling interval to query price of 1 hour, querying for a period of a week. 
Let us also call \textit{base price} the minimum fare charged for a route by the standard Uber pricing (UberX in NYC is $\$2.15 mile + 40 cents/minute$).

There are a few key observations to be highlighted here. First, the price value of 
a single route can vary significantly over time. Considering a $168$-hour window of observation (1 week), 
routes may surge frequently, typically three or four times a day, with surge periods lastings sometimes
a few hours. Sometimes route prices can 
increase significantly in absolute value, an increase than can even be in the order of tens of U.S. Dollars, with respect to the 
minimum \textit{base price}. 
Second, the temporal dynamics of the route prices appear to be correlated, but not automatically, as one observes many times when some routes surge and others are in the base price. 
% Second, the temporal dynamics of the route prices appear to be correlated.
% However, there are frequent cases of routes that surge when others are offered in the base price. 
This observation is expected, as routes originate from different areas, each characterized by different driver supply and customer demand patterns and, as a consequence, different surge patterns.
% The latter observation is simply an effect of the fact that routes can originate at different areas,
% with those areas featuring different driver supply and customer demand patterns, and as a consequence,
% their surge patterns should be different. 

\begin{figure}
\centering
\includegraphics[width=1.0\linewidth]{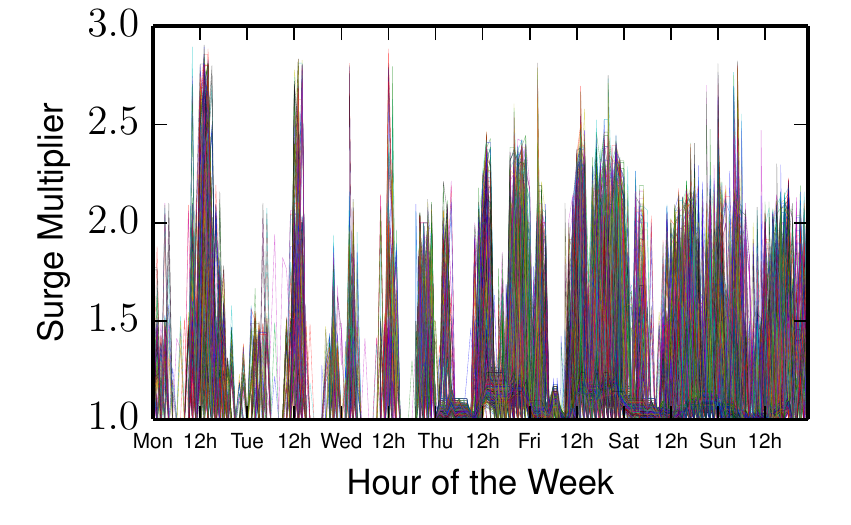}
\caption{Route Surge Multipliers during the course of a week.}  
\label{priceEvolMult}  
\end{figure}

% multiplier
Surge pricing proceeds by multiplying the baseline price depending on offer and demand. For this reason, we show in Figure~\ref{priceEvolMult} how the price multiplier of a route evolves in time.
Formally, we define the \textit{surge multiplier} of a route at time $t$ as $\frac{price(i,t)}{base\_price(i)}$, 
where $price(i,t)$ is the Uber X cost of route $i$ during time $t$ and $base\_price(i)$ its base price.

A value of $1$ indicates a base price. One observes several spikes on the curves representing the different routes, with the frequent presence of large multiplier values. This pattern confirms the observations made in Figure~\ref{priceEvol}. Note that in the window of observation (a weekly time window in May 2015) and for the routes considered for this experiment, the multipliers are capped under a $\times 3$ multiplier. This cap is the reflect of the price control designed by the company. While capping is a common practice in many modern transportation systems~\cite{capping}, in the case of Uber it seems to be a company induced policy, and not an external control applied by local regulatory authorities. Capping in this case may have been enabled due to cases of extreme charges on Uber customers reported publicly in the past~\cite{fiasco}.

% \begin{figure}
% \centering
% \includegraphics[width=1.0\linewidth]{images/surgePricedTrips.pdf}
% \caption{Surge Priced Trips}  
% \label{queryDistr}  
% \end{figure} 

%%%%%%%% MEASURING SURGE PRICING  %%%%%%%%%%%

So far, the most counter intuitive observation regarding Uber's pricing tactics, is that surge is not a rare event. 
While we have no measure of how many journeys are actually purchased through Uber at a surge price, we can exploit the usage statistics of our app, alongside the surge patterns of the corresponding routes 
to provide an estimate. To do so, we exploit the usage frequency statistics shown in Figure~\ref{queryWeekly}. The frequency of user queries is a proxy to the trips purchased in a given hour, noted here as $P_t$ for hours $t \in 1 \dots T$, where $T=168$. For a given route $i$, we note whether at a given hour $t$, it has 
been on surge or not. For example, given a route $i$ and an hour $t$, we can generate a time series $S$ of binary values $s_
{it}$, where $s_{it} = 1$ if the route is priced at surge in that hour, or $s_{it} = 0$
otherwise. Through a simple multiplication of the two time series $P$ and $S$, considering the set of all $N$ routes, we can estimate the fraction of trips purchased at surge, $ST$, in the following manner:
\begin{equation}
ST = \frac{\sum_{i=1}^{N} \sum_{t=1}^{T} s_{it}\times P_t}{N \times \sum_{t=1}^{T} P_t}
\end{equation}

Considering a sample of $800$ routes in New York City and pricing data from a week in May 2015, we have noted 
that more than $1$ in $4$ Uber X trips are purchased at a price higher than the standard \textit{base price}. Of course, this is an indicative figure and corresponds to a simplification of a complex reality. 
The main assumption is that the time evolution of the number of trips purchased, modeled by $P$, is the same over different areas in the city.  
Further, numbers may vary across different time windows either because the supply-demand balance drifts over time, or because Uber changes its surge pricing algorithm.

%%%%%%%% CONTROLLED EXPERIMENT %%%%%%%%%%%
\paragraph*{\textbf{A Surge Pricing Experiment}}
The observations made in the previous section are instructive, but they do not provide an explanation for the underlying mechanics driving surge pricing. As discussed in Section~\ref{sec:analysis}, Uber’s pricing model is known to be based on supply and demand balance~\cite{uberPricing}. It is unclear, however, if demand is evaluated only at origin, or instead if a more complex recipe, incorporating perhaps the overall demand dynamics in the city, is considered.

\begin{figure}
\centering
\includegraphics[width=1.0\linewidth]{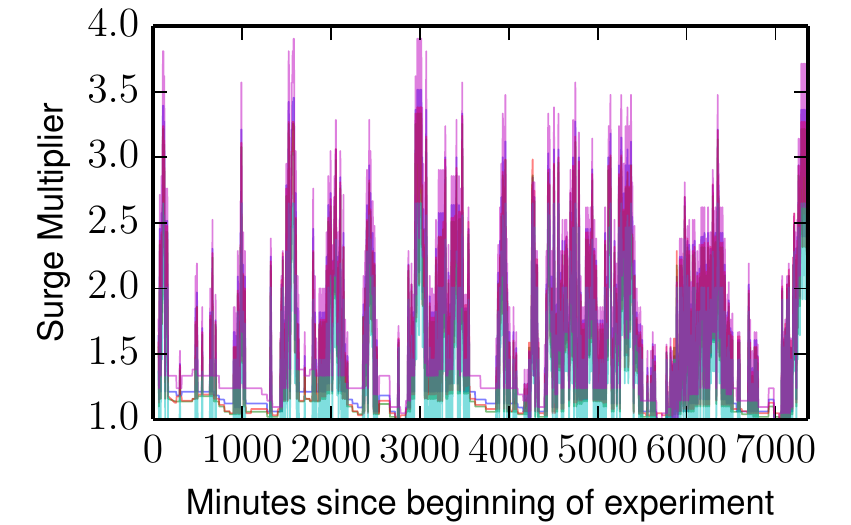}
\caption{Surge Experiment where we control for the origin point (set to Times Square). 
The temporal evolution of surge multipliers is noted for five routes leaving the origin.}  
\label{controlO}  
\end{figure}

In order to understand this mechanism, we perform the following experiment.
For a given origin $O$ in the center of New York (Times Square) we query the Uber API for routes that originate in $O$, and ending in different geographic endpoints sampled randomly. 
If surge pricing was to depend only on demand, the tested routes would be in pure temporal synchronicity.
In Figure~\ref{controlO}
we show the price evolution of a sample of $5$ routes. Our queries were performed at a high frequency of  $\frac{1}{25}$ queries/sec, to allow for the collection of finer time series. 
The results demonstrate that surge pricing strongly depends on the origin point.
Considering all possible pairs of routes we have measured a mean correlation between their time
series equal to $Pearson's~r=0.96$.
Despite the correlation of prices across time, however, we have also observed minor discrepancies. Those could be due to either delays in server responses from Uber's API, or instead to other factors, for instance variations in demand in other regions of the city.

\begin{figure}
\centering
\includegraphics[width=1.0\linewidth]{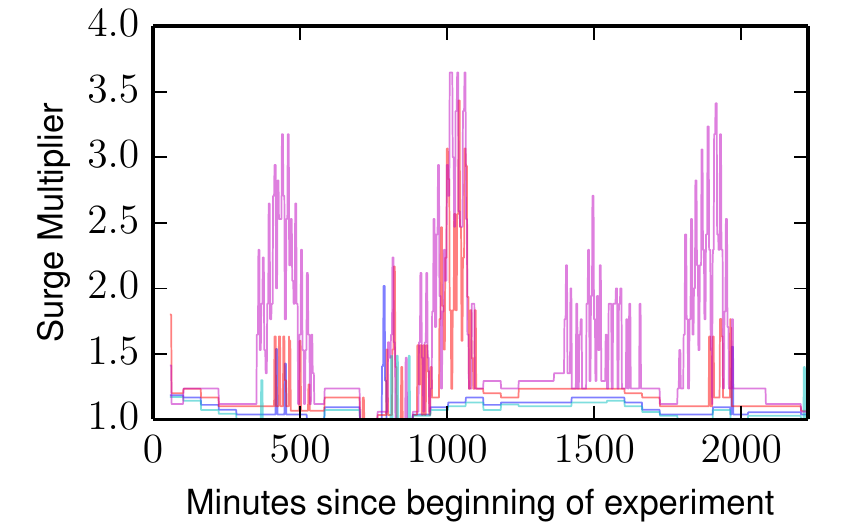}
\caption{Surge Experiment where we control for the destination point. 
The temporal evolution of surge multipliers is noted for five routes reaching the destination (Times Square).}  
\label{controlD}  
\end{figure}

To test the latter hypothesis, we perform a similar experiment but with the control point reversed. That is we test variations in prices among routes that start at different origin points $O$, but 
end at the same destination $D$. In Figure~\ref{controlD} we observe that the price evolution also present correlations,
but to a lesser extent than those of Figure~\ref{controlO}.
In this case, the mean correlation value between all time series pairs was equal to a $Pearson's~r=0.57$.
This result is either due to the existence of spatial correlations of offer and demand across the city, or to the incorporation of data at the destination in order to determine the price of a trip. From an economic perspective, the latter hypothesis is understandable, as Uber would benefit from having their drivers move to areas with a high demand.  

%%%%%%%% SURGE PREDICTION  %%%%%%%%%%%
\paragraph*{\textbf{Geographic Hierachy of Surge Pricing}}
Surge pricing depends on variations of the service’s demand on the side of users and supply on the side of the drivers. Uber’s application permissions allows for access to location information about their users in real time, and it is thus likely that their model to estimate is based on this information. In addition, it is well-known  in the urban research literature that population density exhibits heterogeneous geographic distribution patterns~\cite{bak1997nature}, typically reflecting a more densely populated urban core and a more sparsely populated periphery. 

\begin{figure}
\centering
\includegraphics[width=1.0\linewidth]{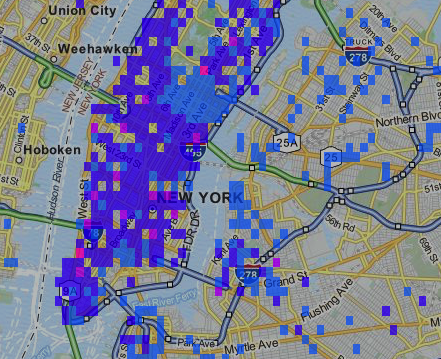}
\caption{Area Surge Geographic Heatmap for different geographic areas in New York.}  
\label{heatmapSurge}  
\end{figure}

In this context, predicting the exact time series of route prices may be a challenging prediction task. Yet, if we assume that different areas in the city are characterised by different population densities, user demand is expected to be distributed similarly. We explore this possibility in Figure~\ref{heatmapSurge}, where we visualize the spatial distribution of surge pricing multipliers over different areas in the city where the users of our application have travelled. 
Formally the \textit{average surge multiplier} of a route $i$ as the mean of all its price evaluations over time:
\begin{equation}
AverageSurgeMultiplier =\sum_{t=1}^{T} \frac{\frac{price(i,t)}{base\_price(i)}}{T}
\end{equation}
Then the mean surge of an area is measured by taking into account the $AverageSurgeMultiplier$ values for all routes that leave a given cell area (i.e., the cell is origin for these routes). 

\begin{figure}
\centering
\includegraphics[width=1.0\linewidth]{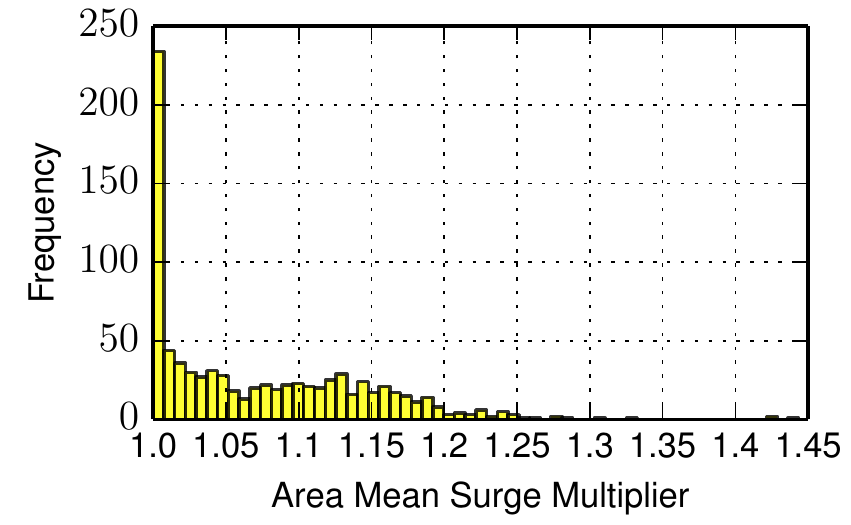}
\caption{Distribution of mean surge multiplier values for the $840$ cell areas in New York. 
The mean surge multiplier is measured considering all surge multipliers of the routes that
have an area as their origin point.}  
\label{surgeacrossareas}  
\end{figure}

A visual inspection supports the idea that indeed more central and dense areas are more prone to surge, associated to a higher average multiplier. An analytical viewpoint on the distribution of the numerical values of mean area surge is provided through Figure~\ref{surgeacrossareas} where a frequency histogram is shown. Most areas in the periphery of the city have an average
surge multiplier equal to $1.0$, but there is a considerable percentage, almost $70\%$ which has a higher multiplier. Our goal next is to predict those areas that are more likely to be prone to surge pricing.

\paragraph*{\textbf{Predicting Surge Pricing}}
Finally, we investigate whether demand can be estimated by combining different datasets, without using Uber information subject to API limitations. In particular, our aim is to predict surge multipliers in different areas, and therefore the surge hierarchy in urban neighborhoods in New York. We reduce this problem to a ranking task where our goal is to rank areas from higher to lower surge values. To do so, we need to estimate local demand and local offer, but we will only focus on the former, as we have no information about the residence of Uber drivers nor about their whereabouts. For this reason, we make the assumption that driver supply is uniform in the city. 

To estimate demand, we combine two different datasets. First, we use the yellow taxi dataset described above, where the number of trips per geographic area can be recorded. The yellow taxi user base of course may not be the same as Uber's, but given the competition between the two companies, an overlap is expected. Secondly, we import a dataset from Foursquare and in particular the venues and check-ins of the location-based service in New York city during $2011$.
This data provides us estimates of urban place and population density but also the number of transportation hubs, as the latter are expected to be popular destinations for taxis. The dataset signals are combined with a supervised learning model, that is a Decision Tree Regressor~\cite{quinlan1992learning}, where we have set a maximum tree depth equal to $20$ and trained and tested using the Leave-One-Out Error~\cite{hastie2009elements} technique.
\\
\textbf{Results:}
In Table~\ref{TablePearson}, we present the Pearson correlation r between the average surge pricing multiplier observed in the $840$ areas visualised in Figure~\ref{heatmapSurge}, the four datasets used to estimate Uber X demand, and the supervised learning model. Among individual signals, the correlation is highest with the frequency of yellow cab trips ($r= 0.43$). The number of Foursquare Places is second with a score $r=0.42$. However, the best score is, by far, obtained with the Decision Tree ($r=0.82$). This result is impressive given that we measure correlations between variables collected from distinct technological systems. Note also that despite its low correlation ($r=0.07$), the incorporation of the frequency of Foursquare travel spots as a feature in the supervised learning model has helped to improve performance from $r=0.78$ to $r=0.82$. 

\begin{table}
\caption{Pearson's r correlation }
\centering
\begin{tabular}{|cc|}
\hline
\textbf{Feature} & \textbf{Pearson's r}\\
\hline
\hline
\centering
Yellow Cab Trips & $0.43$ \\ \hline
Foursquare Places & $0.42$ \\ \hline
Foursquare Check-ins & $0.35$ \\ \hline
Foursquare Travel Spots & $0.07$ \\ \hline
Decision Tree Regressor & $0.82$ \\ 
\hline
\end{tabular}
\label{TablePearson}
\end{table}

Finally, we define a ranking task, aiming at ranking areas based on their average surge price. The quality of a ranking is measured in terms of $NDCG$ metric, well-known in information retrieval theory. Three out of four individual signals achieve an $NDCG@100$ score in the range $0.87-0.89$, with the number of Foursquare Travel Spots scoring $0.84$. Note that a random baseline (ranks areas by shuffling randomly the list of areas) achieves a score of $0.83$. As in the case of the Pearson correlation metric r, the Decision Tree model outperforms individual models, attaining an NDCG score of $0.97$. 

\begin{table}
\caption{NDCG Scores for the Ranking Task}
\centering
\begin{tabular}{|cc|}
\hline
\textbf{Feature} & \textbf{NDCG@100}\\
\hline
\hline
\centering
Yellow Cab Trips & $0.87$ \\ \hline
Foursquare Places & $0.89$ \\ \hline
Foursquare Check-ins & $0.88$ \\ \hline
Foursquare Travel Spots & $0.84$ \\ \hline
Decision Tree Regressor & $0.97$ \\ \hline
Random Baseline & $0.83$ \\
\hline
\end{tabular}
\label{TablePearson}
\end{table}

%DESCRIBE RESULTS %

% \begin{figure}
% \centering
% \includegraphics[width=1.0\linewidth]{images/heatmapYellow.png}
% \caption{Yellow Heatmap}  
% \label{heatmapYellow}  
% \end{figure}

% \begin{figure}
% \centering
% \includegraphics[width=1.0\linewidth]{images/heatmapFoursquare.png}
% \caption{Foursquare Place Density Heatmap}  
% \label{heatmapFoursquare}  
% \end{figure}

\section{related works}
\label{sec:related}
This paper is at the border between several disciplines related to urban data science, including urban data mining, spatial economics and mobility studies on taxi datasets. Urban data mining has been gaining traction in recent years due to the increasing availability of datasets, and to strategic decisions of many urban authorities to realize the vision of smart cities. Related to this work, a popular idea is to analyze activity in urban transportation systems to estimate commuter costs and propose data mining methods to reduce them~\cite{sobolevsky2015cities, lathia2011mining, LathiaSmart}.
Mining data becoming publically available through sharing bicycle transportation schemes 
has been another common line of research~\cite{lathia2012measuring, froehlich2008measuring, schuijbroek2013inventory}. More generally, data from social media has been mined to digitally
represent and model various aspects of urban reality~\cite{Quercia2014}, whereas telecom  and location-based services data for urban activity recognition~\cite{Soto2011, noulas2013exploiting}. 

Related in terms of data sources, let us also mention efforts to mine spatial trajectories of taxi mobility in the field of urban computing~\cite{yuan2010t, zheng2011computing, zheng2011urban}. The dataset of Yellow Cabs studied in the present work has been exploited recently to quantify the benefits of vehicle pooling in urban environments~\cite{santi2014quantifying}. To the best of our knowledge, however, a combination of mobility data with financial information, as considered here, is novel, as is the idea to develop data mining solutions for transparency in urban taxi transport. Our hope is that similar works will follow as more and more datasets become available, with a potential benefit not only to urban transport, but also in the field of spatial economics in general~\cite{beckmann1985spatial, hale2003location}. In this direction, data mining techniques have recently been applied to identify ideal locations to set up new retail facilities in cities~\cite{karamshuk2013geo}.

\section{conlusion and future work}
\label{sec:discussion}
The findings of the present work have great implications both for the future of urban
transport, but also for data mining research.

First, as new technologies disrupt traditionally established sectors new norms are likely
to emerge. As we have seen the case of Uber has dramatically altered the economic landscape 
of transport by taxi. While our work has focused on the example case of New York, similar
trends are being observed in other metropolitan environments where Uber like services 
launch. Regarding this evolution, in Section~\ref{sec:analysis}, we have demonstrated 
how modern open datasets that describe urban transport can help towards a more transparent
economic reality in a sector that now experiences massive changes. Moreover, these datasets can be exploited by mobile applications (Section~\ref{sec:app}) that have the potential to reach
thousands of users and help obtain significant savings during their daily commutes as we have shown
in Section~\ref{sec:evaluation}. 

Secondly, we have seen that is possible to exploit observed data in order to reverse engineer, to some
extent, the functionality of complex algorithms that are deployed in the real world by technology 
companies. As these disruptions continue so does the need for work in the emerging field of algorithmic
transparency~\cite{diakopoulos2014algorithmic} emerges. Having focused on Uber's popular surge pricing
methods, we have shown it presents tractable characteristics which are mainly sourced in local demand 
patterns posed by mobile users. Interestingly, as we have shown in Section~\ref{sec:surge}, it is possible to estimate average demand at an area, and therefore surge, using exogenous to Uber data. The geographic
characterization of surge we have performed can be incorporated in our application, or similar ones, 
to improve user experience and help them save more. For example, consultation on how long they need to wait, or which block they need to walk into for calling a taxi, could help them avoid surge pricing.

Overall, we believe that these observations can inspire novel work in the field of data mining. 
The idea of incorporating datasets from multiple services (Uber, Foursquare, Yellow Cabs) for innovative applications as we have done in the present work corresponds to a new frontier in the areas
of big data mining and machine learning. Further, while we have performed a geographic prediction of surge~\ref{sec:surge}, new approaches could be developed that identify the evolution of surge dynamically over time. In this context, the development of algorithms and models
that realize the spatio-temporal dynamics of complex urban systems using modern datasets from 
multiple location-based services or transport systems could be an interesting future direction
to consider.

% \section{conclusion}
% \label{sec:conclusion}
% \input{conclusion}

% \section{Reverse Engineering Surge Pricing Algorithmics}

% \begin{figure}
% \centering
% \includegraphics[width=0.7\linewidth]{images/priceEvolution.pdf}
% \caption{Price Evolution}  
% \label{queryDistr}  
% \end{figure} 

% \begin{figure}
% \centering
% \includegraphics[width=0.7\linewidth]{images/priceEvolutionDiffsMult.pdf}
% \caption{Price Evolution Mult}  
% \label{queryDistr}  
% \end{figure}

% \begin{figure}
% \centering
% \includegraphics[width=0.7\linewidth]{images/priceEvolutionControlledExpNYRat.pdf}
% \caption{Surge Experiement Control Origin}  
% \label{queryDistr}  
% \end{figure}

% \begin{figure}
% \centering
% \includegraphics[width=0.7\linewidth]{images/priceEvolutionControlledExpNYRat_rev.pdf}
% \caption{Surge Experiement Control Destination}  
% \label{queryDistr}  
% \end{figure}

% \begin{figure}
% \centering
% \includegraphics[width=0.7\linewidth]{images/surgePricedTrips.pdf}
% \caption{Surge Priced Trips}  
% \label{queryDistr}  
% \end{figure} 

% \section{Related Work}

\small
\bibliographystyle{plain}
\bibliography{biblio}
\end{document}